\documentclass[pdflatex,sn-vancouver-num]{sn-jnl}

\usepackage{graphicx}%
\usepackage{multirow}%
\usepackage{amsmath,amssymb,amsfonts}%
\usepackage{amsthm}%
\usepackage{mathrsfs}%
\usepackage[title]{appendix}%
\usepackage{xcolor}%
\usepackage{textcomp}%
\usepackage{manyfoot}%
\usepackage{booktabs}%
\usepackage{algorithm}%
\usepackage{algorithmicx}%
\usepackage{algpseudocode}%
\usepackage{listings}%
\usepackage{hyperref}
\usepackage{bookmark}

\theoremstyle{thmstyleone}%
\theoremstyle{thmstyletwo}%

\theoremstyle{thmstylethree}%

\begin{document}

\title{The temporal resolution limit in quantum sensing}

\author[1,2]{\fnm{Cong-Gang} \sur{Song}}
\author*[3]{\fnm{Qing-yu} \sur{Cai}}\email{qycai@hainanu.edu.cn}
\affil[1]{Innovation Academy for Precision Measurement Science and Technology, Chinese Academy of Sciences, Wuhan 430071, China}
\affil[2]{University of Chinese Academy of Sciences, Beijing 100049, China}
\affil[3]{Center for Theoretical Physics, School of Physics and Optoelectronic Engineering, Hainan University, Haikou, 570228, China}

\abstract{Temporal resolution is a critical figure of merit in quantum sensing. This study combines the distinguishable condition of quantum states with quantum speed limits to establish a lower bound on interrogation time. When the interrogation time falls below this bound, the output state becomes statistically indistinguishable from the input state, and the information will inevitably be lost in noise. Without loss of generality, we extend these conclusions to time-dependent signal Hamiltonian. In theory, leveraging certain quantum control techniques allows us to calculate the minimum interrogation time for arbitrary signal Hamiltonian. Finally, we illustrate the impact of quantum speed limits on magnetic field measurements and temporal resolution.}





\maketitle

\section{Introduction}

In quantum metrology, detecting a weak signal is equivalent to distinguishing two neighboring quantum states~\cite{RN271}.
Quantum sensing comprises three fundamental stages: the preparation of a known input quantum state $|\psi\rangle$, the interaction between the input state and signal, and the readout of the unknown output state $\left|\psi^{\prime}\right\rangle=e^{-i \varphi \hat{H}}|\psi\rangle$.
The phase $\varphi$ encapsulates the signal's information, thus the primary objective of quantum sensing is to encode and decode this phase information, maximizing the distinction between input and output states to extract signal information. The boundary of measurement precision represents a pivotal challenge in quantum sensing. To date, the most prevalent benchmark is the quantum Cram\'er-Rao bound (QCRB) derived from quantum Fisher information~\cite{RN811,RN271}. Optimal precision can attain the standard quantum limit of $1/\sqrt{n}$ utilizing independent quantum states for measurement~\cite{RN308, RN272, RN447,RN295, RN282}. The Heisenberg limit, scaling as $1/n$, is achievable through the utilization of squeezed or entangled states~\cite{RN295, RN282,RN447, RN279}, and may even extend to the ``super-Heisenberg limit" scaling as $1/n^k$ with $k>1$ when considering many-body interactions or nonlinear effects~\cite{RN293, RN292, RN277, RN280, RN469}. However, temporal resolution, another critical figure of merit in quantum sensing, has not been given comparable emphasis. 

A natural idea is that temporal resolution should be closely related to the quantum speed limits. At present, the two most famous quantum speed limits are the Mandelstam-Tamm bound (MTB) and the Margolus-Levitin bound (MLB)~\cite{RN296,RN256,RN304}.
In fundamental research, quantum speed limits have garnered significant attention in various areas, including open systems ~\cite{RN359,RN360,RN361,RN314,RN390}, the quantum-to-classical transition ~\cite{RN311,RN366}, and the dynamics of time-dependent systems ~\cite{RN359,RN388,RN412,RN410}. In practical applications, these quantum limits have been widely employed in fields such as quantum computing~\cite{RN326,RN460}, and quantum optimal control~\cite{RN393}. Within the realm of quantum sensing, some studies focus on the relationship between quantum speed limits and quantum Fisher information~\cite{RN361,RN390,RN291,RN370,RN401,RN402}, while others attempt to derive precision bounds directly from quantum speed limits \cite{RN297,RN922}. However, for a long time, little research has directly investigated the temporal resolution limit in quantum sensing from the perspective of quantum speed limits except for a recent work \cite{RN977}. This scarcity of research might stem from the fact that in most scenarios, the signals considered are static, rendering temporal resolution less important. In such cases, longer interrogation time typically lead to higher measurement precision, prompting researchers to prioritize the upper bound of interrogation time rather than its lower bound. However, in decoherent environments or certain noisy measurement schemes, longer interrogation time does not necessarily lead to better results. In some cases, the optimal interrogation time may even approach zero \cite{RN537,RN533}. A natural question arises: can the interrogation time be reduced to an arbitrarily small value? On the other hand, when signals are time-dependent, shorter interrogation times enable the capture of finer signal details, making temporal resolution critically important in such contexts.

The issue of temporal resolution has recently garnered increasing attention in theoretical and experimental works. A notable example is a recent theoretical work \cite{RN977} that proposed applying both control sequences and the signal Hamiltonian concurrently to the quantum state.  This method circumvents the need for dedicated interrogation time, enabling temporal resolution to be dictated solely by the control pulse power and potentially achieving the quantum speed limit. Subsequent experimental work \cite{RN1055} quickly verified this concept. Similarly, another experiment enhanced temporal resolution by generating electron wave packets with shorter temporal widths \cite{RN1054}. The main idea of these schemes is to shorten the control pulse or electron wave packet to minimize the interaction time between the quantum state and the signal, thereby obtaining a sufficiently high temporal resolution, but this will also affect the sensitivity and signal-to-noise ratio of the signal. Therefore, theoretically,  the time resolution cannot be infinitely improved under limited resources, otherwise the signal will be submerged in the noise and difficult to distinguish.

In this paper, we investigate the fundamental lower bound of temporal resolution. To this end, we focus on the signal–state interaction process, i.e., the interrogation stage. Starting from the quantum state distinguishable condition, we derive an expression for the minimum detectable signal in terms of fidelity, and then combined with the quantum speed limits, we obtained the lower bound of the time resolution. Under ideal conditions, this bound can be further reduced by increasing available resources—for example, by increasing the number of repeated measurements, using more quantum states, or employing entangled states—thereby improving the minimum detectable signal or accelerating quantum evolution. To clarify these effects, we discussed in detail the impact of various measurement resources on the quantum speed limits.
Given the lack of prior information, the quantum speed limit may not always be saturated. We analyze the applicability of two quantum speed limits at different levels of prior knowledge. In practical quantum sensing, the signal is often time-dependent, which makes the temporal lower limit in time-dependent scenarios a more interesting question. With the help of quantum control techniques, the quantum states can be kept in the optimal detection state throughout the measurement process~\cite{RN280,RN405,RN324}. This allows the direct calculation of the quantum speed limit and time resolution of arbitrary Hamiltonian. Finally, we illustrate the impact of the quantum speed limit on magnetic field measurements and their associated time resolution through two specific examples.

\section{Distinguishable condition of quantum states}
\label{DCOQS}

Distinguishing between two neighboring quantum states, denoted as $|\psi\rangle$ and $\left|\psi^{\prime}\right\rangle$, is inherently challenging,  as only orthogonal states are fully distinguishable. However, by repeatedly measuring and comparing the differences in the probability distributions, it is statistically feasible to distinguish between two quantum states. According to Wootters' proposition \cite{RN301}, two adjacent quantum states are distinguishable if the following condition is satisfied:
\begin{equation} \label{SNR}
	\left|p-p^{\prime}\right| \geq \Delta p+\Delta p^{\prime},
\end{equation}
where $p$, $p^{\prime}$ are the probability estimators derived from the measurement results, and $\Delta p$, $\Delta p^{\prime}$ are the corresponding uncertainties (usually measured in standard deviation).

For two-level quantum states, the results of projection measurement meet the binomial distribution. According to the knowledge of probability theory \cite{RN301,RN149,RN71}, it can be obtained
\begin{equation}
	\Delta p=\left[\frac{p(1-p)}{N}\right]^{\frac{1}{2}},
\end{equation}
where $N$ is the number of independent repeated measurement. As $N$ increases, the corresponding uncertainty decreases, enhancing the distinguishability of two quantum states. There are two straightforward methods to increase $N$.
For instance, one can perform $N$ repeated measurements on a single quantum state or conduct a single measurement on $N$ independent quantum states. These two approaches achieve equivalent outcomes. The difference is that the former consumes temporal resources, thereby impacting time resolution, while the latter consumes spatial resources, thus affecting spatial resolution. 

In essence, we interpret the discrepancy between probability distributions as the signal and the aggregate of uncertainties as noise. Consequently, the significance of Eq. (\ref{SNR}) lies in asserting that the signal-to-noise ratio (SNR) equals or exceeds one (i.e.,SNR$ \geq 1$). In other words, effective differentiation between two probability distributions (quantum states) is achievable only when the signal surpasses the noise. Otherwise, the signal information becomes submerged in projection noise, rendering the two probability distributions indistinguishable. Therefore, Eq. (\ref{SNR}) can be defined as the distinguishable condition for probability distributions, applicable to both classical and quantum scenarios.

In contrast to classical probability distributions, quantum states offer non-uniqueness in their probability distributions due to the availability of various measurement bases. The discrimination ability of different measurement bases is also different. 
For instance, consider two orthogonal quantum states: $|+\rangle=(|0\rangle+|1\rangle)/ \sqrt{2}$ and $|-\rangle=(|0\rangle-|1\rangle)/ \sqrt{2}$. When selecting $|0\rangle$ or $|1\rangle$ as the measurement basis, regardless of the augmentation of measurement resources $n$, discrimination between $|+\rangle$ and $|-\rangle$ remains unattainable. Conversely, opting for $|+\rangle$ or $|-\rangle$ as the basis vector enables their differentiation with just one measurement.

To maximize the distinguishability between two neighboring quantum states, directly selecting one of them as the measurement basis is optimal \cite{RN301,tradeoff}. In this case,
\begin{equation}
	\begin{aligned}
		p(|\psi\rangle)&=|\langle\psi \mid \psi\rangle|^{2}=F(|\psi\rangle,|\psi\rangle)=1, \\
		p\left(\left|\psi^{\prime}\right\rangle\right)&=\left|\left\langle\psi \mid \psi^{\prime}\right\rangle\right|^{2}=F\left(|\psi\rangle,\left|\psi^{\prime}\right\rangle\right).
	\end{aligned}
\end{equation}
The probability of the measurement result is the fidelity $F$ between the quantum state and the measurement basis\cite{RN1056}. Therefore, the distinguishable condition of probability distributions Eq. (\ref{SNR}) can be written as
\begin{equation} \label{1-F}
	|1-F| \geq \Delta F.
\end{equation} 
For $N$ independent measurements, $\Delta F=\sqrt{F(1-F)/N}$, it can be simplified as
\begin{equation} \label{FFODC}
	F \leq \frac{N}{N+1},
\end{equation}
which is the fidelity expression of the distinguishable condition of quantum states.
Eq.(\ref{FFODC}) is universal and independent of any specific schemes, and
the minimum detectable signal (i.e., precision $\delta \varphi$) implied within this inequality.
For a weak signal, $\delta \varphi  \propto [NF_Q(\varphi)]^{-1/2}$ \cite{tradeoff}, which is consistent with the QCRB based on quantum Fisher information $F_Q(\varphi)$.

\section{Quantum speed limits for time-dependent Hamiltonian }
\label{QSL-TDH}

For a finite measurement resource $N$, the distinguishable maximum critical fidelity is $F_0=N/(N+1)$.  This implies that the fidelity between the input and output states--driven solely by the signal field--must exceed the critical value $F_0$ for the signal to be considered effectively measured. The required evolution time corresponds to the minimum interrogation time, which can be well-estimated using the quantum speed limits.

For the time-independent Hamiltonian $H$ and the corresponding unitary evolution $\left|\psi_t\right\rangle=e^{-i H t/\hbar}|\psi_0\rangle$, two common quantum speed limits are defined for a given fidelity $F=\left|\left\langle\psi_t \mid \psi_0\right\rangle\right|^{2}$, that is the Mandelstam-Tamm bound and the Margolus-Levitin bound ~\cite{RN296,RN256,RN304}
\begin{equation} \label{QSLS}
	\begin{gathered}
		\tau_{ML} = \frac{\pi \hbar}{2\langle H-E_g\rangle} \cdot \alpha(F), \\
		\tau_{MT} = \frac{\pi \hbar}{2 \Delta H} \cdot \beta(F),
	\end{gathered}
\end{equation}
where $\beta(F)=2 \arccos \sqrt{F}/\pi$, $\alpha(F) \simeq \beta^{2}(F)=4\arccos ^{2} \sqrt{F}/\pi^{2}$, and $0 \leq \alpha(F),\beta(F) \leq 1$, which quantify the distance between the input and output states. It can also see the appendix \ref{APPB}, \ref{APPC} for relevant derivations. To satisfy the quantum state distinguishable condition (i.e., Eq.(\ref{FFODC})),  the actual evolution time must fulfill $t\geq max(\tau_{ML},\tau_{MT})$; otherwise, the signal will be submerged in the noise.
By replacing $t$ with the parameter $\varphi$, and replacing the Hamiltonian $H$ with the signal generator $G$, these two quantum speed limits directly provide a lower bound on the precision of parameter estimation. For large $N$, higher-order terms can be neglected, so $\arccos\sqrt{F_0} \approx 1/\sqrt{N}$. Substituting this result into Eq. (\ref{QSLS}), we can get
\begin{equation}
	\delta\varphi \geq max(\frac{2\hbar/\pi}{ N\langle G-G_g\rangle},\frac{\hbar}{\sqrt{N} \Delta G}).
\end{equation}
Similar forms of precision limit can also be found in the references \cite{tradeoff,RN277,RN297}.

\begin{figure*}[!hbt]
	\centering
	\includegraphics[width=14cm]{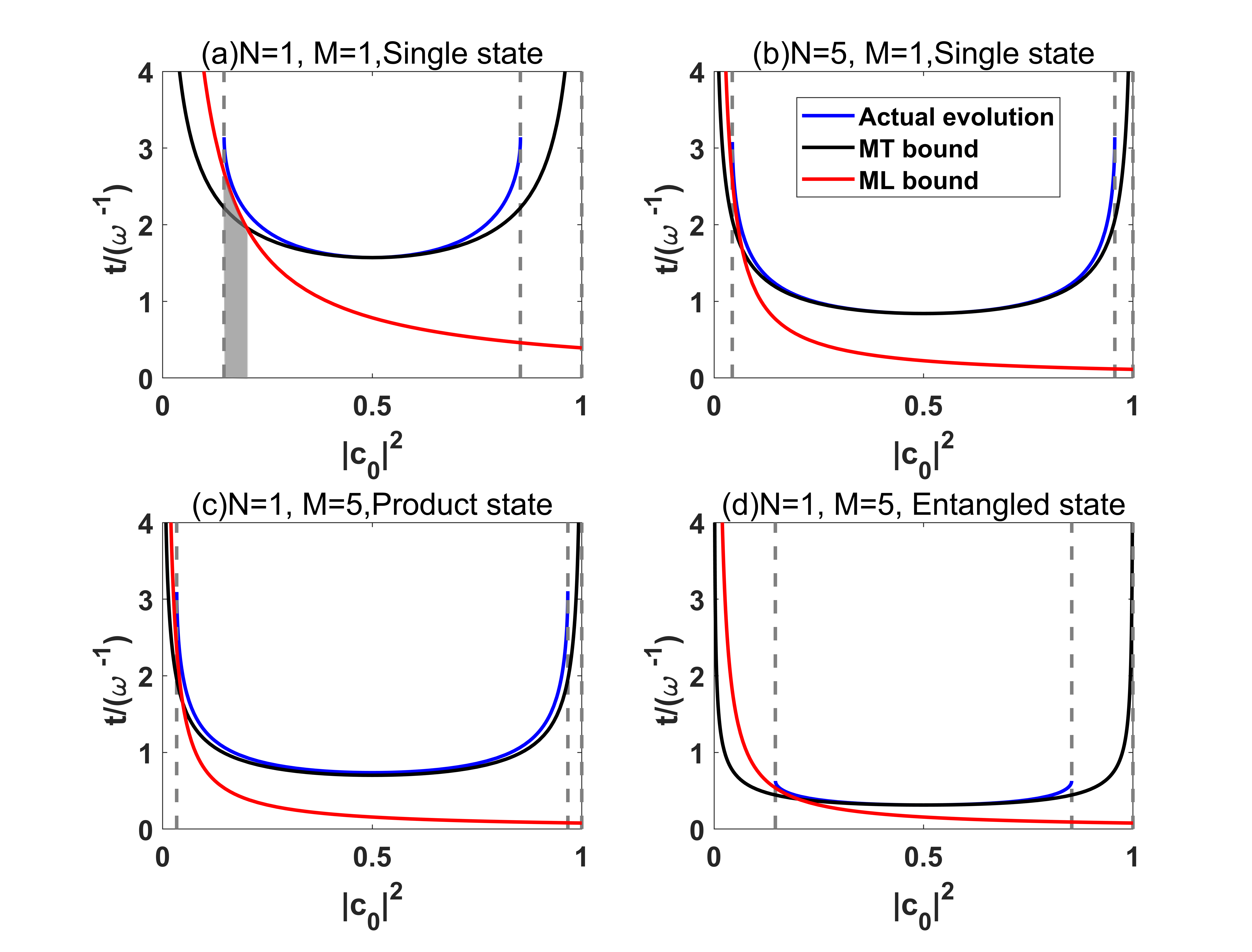}
	\caption{Quantum speed limits and actual evolution time. Different $|c_0|^2$ correspond to different directions $\vec{S}$, which is unknown in general. The equal-weight superposition state (i.e.$|c_0|^2=0.5$) is optimal, minimizing the actual evolution time (blue line) and yielding the tightest MTB. As $|c_0|^2$ gradually decreases to a certain range (e.g. the gray area in (a)), MLB (red line) can provide better estimation compared to MTB (black line). When $|c_0|^2$ is close to 0 or 1,  the critical fidelity cannot be attained, preventing effective measurement of the signal, which corresponds to the range outside the dotted line. We further explored the impact of various resources on  the quantum speed limits and the actual evolution time. $N$ represents the number of independent repeated measurements, which is used to reduce statistical noise so that smaller signal can be detected, thereby improving time resolution, as shown in (b). $M$ represents the number of qubits contained in the probe state, which is used to speed up the evolution so that more signal information can be picked up in a limited time. For example, as demonstrated in (c), $M$-body product states are used to improve time resolution. If $M$-body GHZ entangled states are used, the evolution speed can be further accelerated, as shown in (d). }\label{fig-QSL}
\end{figure*}

The results above can be extended to multiple quantum states. For measurements on $M$-body separable states, probed as a product state, we have $\Delta \hat{H}^{pro}=\sqrt{M}\Delta \hat{H} ,
\langle \hat{H}\rangle^{pro}=M\langle\hat{H}\rangle $.
For $M$-body GHZ entangled states, $\Delta \hat{H}^{ent}=M\Delta \hat{H} ,
\langle \hat{H}\rangle^{ent}=M\langle\hat{H}\rangle $.
Consequently, the minimum time is given by $\tau_{MT}=\sqrt{M}\cdot \tau _{MT}^{pro}  =M \cdot \tau _{MT}^{ent}$,
$\tau_{ML}=M\cdot \tau _{ML}^{pro}  =M \cdot \tau _{ML}^{ent}$. In other words, many-body quantum states can significantly reduce the quantum speed limit, which in essence speeds up the evolution, as shown in Fig.\ref{fig-QSL}(c)(d).
On the other hand, using multiple separable states for $N$ repeated measurements can reduce statistical noise and thereby increase the critical fidelity $F_0$, but it does not affect the evolution speed.
In this case, the minimum evolution time also decreases, however, unlike the former, the minimum time decreases because the distances $\alpha(F)$, $\beta(F)$ decrease as $F_0$ increases, as shown in Fig.\ref{fig-QSL}(b).

For an unknown time-independent signal Hamiltonian,  it can be written as $\hat{H}(\omega)= \vec{\sigma} \cdot \vec{S}(\omega)$, where $\vec{\sigma} =(\sigma_x,\sigma_y,\sigma_z)$, represents the direction of the signal Hamiltonian on the Bloch sphere. $\left |\vec{S}  \right |=\hbar \omega/2 $ represents the magnitude of the signal, its eigenvectors are denoted as $\left | \tilde{0} \right \rangle,  \left | \tilde{1}   \right \rangle $, so that a known input state can be written as $\left | \psi  \right \rangle =c_0\left | \tilde{0} \right \rangle+ \sqrt{(1-c_0^2)}\left | \tilde{1}   \right \rangle$. In general, $\vec{S}$ is unknown and hence $c_0$ is also unknown. In Fig.\ref{fig-QSL}, we calculate the quantum speed limits for different $c_0$. It can be found that in most cases MTB provides a better estimation compared to MLB. However, when $|c_0|^2$ is small (e.g. the gray area in the Fig.\ref{fig-QSL}(a)), the MLB yields a better estimation, despite the measurement not being optimal and the estimation time not being saturated. It is noteworthy that the quantum speed limit is saturated only when $|c_0|^2=0.5$,  when the actual evolution time is the shortest. In fact, if sufficient prior information about the direction of the signal $\hat{H}(\omega)$ is available, it is always possible to choose an appropriate basis such that $|c_0|^2=0.5$, thereby saturating the quantum speed limits.

In quantum sensing, the signal is often time-dependent, and so is the Hamiltonian. 
However, the original quantum speed limits given in Eq.(\ref{QSLS}) are not entirely applicable to cases involving time-dependent Hamiltonian. They only hold when $[\hat{H}(t), \hat{H}\left(t^{\prime}\right)]=0$. Further details can be found in appendix \ref{APPB}, \ref{APPC}. More general time-dependent speed limits are often more complex and difficult to saturate \cite{RN359}, making them unsuitable for analyzing time resolution.  
To address this challenge, we express the time-dependent Hamiltonian as  $\hat{H}(\omega,t)= \vec{\sigma} \cdot \vec{S}(\omega,t)$, where $\vec{S}(\omega,t)$ describes the trajectory of the Hamiltonian on the Bloch sphere. We discuss the quantum speed limits in two scenarios.
In the first case, $\vec{S}(\omega,t)=r(t)\cdot \vec{h}(\omega)$, and $|\vec{h}(\omega)|=\hbar \omega/2$. Here, the direction of the Hamiltonian on the Bloch sphere remains fixed along one axis, leading to time-independent eigenvectors of the Hamiltonian $\hat{H}$. It is not difficult to verify that $[\hat{H}(t), \hat{H}\left(t^{\prime}\right)]=0$ is satisfied at this time, so it can be directly calculated
\begin{equation}
	\begin{gathered}
		\Delta \hat{H}(t)=\Delta \hat{h} \cdot|r(t)|, \\
		\langle \hat{H}(t)\rangle=\langle\hat{h}\rangle \cdot r(t),
	\end{gathered}
\end{equation}
where $\hat{h}= \vec{\sigma} \cdot \vec{h}(\omega)$.
Therefore, the two quantum speed limits can be written as
\begin{equation} \label{MTTI}
	\begin{gathered}
		\int_{0}^{t_{MT}} \Delta \hat{h}\cdot|r(t)| d t \geq \frac{\pi \hbar}{2} \cdot \beta(F),\\
		\int_{0}^{t_{ML}}[\langle\hat{h}\rangle \cdot r(t)-E_{g}(t)] dt \geq \frac{\pi \hbar}{2} \cdot \alpha(F),
	\end{gathered}
\end{equation}
where $E_{g}(t)=-\hbar \omega \cdot |r(t)|/2$, is the ground state energy at time $t$. The lower bound on the evolution time is the solution of the above inequalitys.

It is straightforward to verify that, similar to the time-independent case, the quantum speed limits provide the tightest estimate, and the actual evolution time is minimized when the input state is the equal-weight superposition of eigenvectors (i.e.$\left | \tilde{0} \right \rangle,  \left | \tilde{1}   \right \rangle $). 
Moreover, it is well established that the uncertainty in parameter estimation is minimized for equal-weight superposition states  (i.e.$|c_0|^2=0.5$) \cite{RN282}. Thus, the minimum evolution time is inherently linked to optimal measurement. This conclusion follows naturally, as both the MTB and quantum Fisher information are determined by the variance of the generators $\Delta \hat{H}$ and achieve their maximum value when the input state is an equal-weight superposition state.

The finding above indicates that as long as the probe state remains stable in the equal-weight superposition, the measurement results will be optimal and the required time will be minimized. However, for a more general $\hat{H}(\omega,t)$, the direction of the Hamiltonian on the Bloch sphere changes over time, leading to $[\hat{H}(t), \hat{H}\left(t^{\prime}\right)] \ne 0$, the eigenvectors of the Hamiltonian $\hat{H}$ is time-dependent. Consequently, the probe state cannot be maintained in the equal-weight superposition state spontaneously.
In this scenario, the conventional quantum speed limit no longer applies, and the quantum Fisher information fails to reach its optimal value. Fortunately, it has been theoretically and experimentally demonstrated that one can ensure the probe state remains in an equal-weight superposition through quantum control techniques \cite{RN280,RN405}. Furthermore, this technique enables the estimation of general Hamiltonian parameter (e.g. the oscillation frequency of the magnetic field)  not just the multiplicative parameter (e.g. the amplitude of the magnetic field) \cite{RN324}.
In order to estimate a time-dependent signal Hamiltonian $\hat{H}(\omega,t)$, a control Hamiltonian $\hat{H}_c(t)$ can be added \cite{RN405}
\begin{equation}
	\hat{H}_c(t)= \sum_{k}f_k(t) | \tilde{\psi}_k(t)  \rangle  \langle \tilde{\psi} _k(t)  | -\hat{H}(\omega_c,t)+i\sum_{k} | \partial _t \tilde{\psi}_k(t)  \rangle  \langle \tilde{\psi}_k(t) |,
\end{equation}
where $\omega_c$ is an a priori estimate of $\omega$, the choice of $f_k$ have some degrees of freedom, here we can take $f_k=0$ for convenience, and $\tilde{\psi}_k(t)$ are the instantaneous eigenvectors of $\partial _{\omega}\hat{H}(\omega,t)$, that is
\begin{equation}
	\partial _{\omega}\hat{H}(\omega,t)  | \tilde{\psi}_k(t)   \rangle  =\mu _k(t) | \tilde{\psi}_k(t)  \rangle.
\end{equation}
When $\omega$ and $\omega_c$ are sufficiently close, the total Hamiltonian can be written as
\begin{equation}
	\hat{H}_{tot}(t)= \partial _{\omega}\hat{H}(\omega,t)|_{\omega=\omega_c}\cdot \delta \omega+i\sum_{k} | \partial _t \tilde{\psi}_k(t)  \rangle  \langle \tilde{\psi}_k(t) |,
\end{equation}
where $\delta \omega=\omega-\omega_c$. The corresponding evolution operator is given by
\begin{equation}
	U(0\to T )|\tilde{\psi}_k(0)\rangle =\exp (-\frac{i \delta \omega}{\hbar }\int_{0}^{T}\mu _k(t)dt)|\tilde{\psi}_k(T)\rangle.
\end{equation}
We can prepare the input state as $|\psi(0)\rangle =\left ( |\tilde{\psi}_{max}(0)\rangle+|\tilde{\psi}_{min}(0)\rangle  \right ) /\sqrt{2}$, the output state after time $T$ is 
\begin{equation}
	|\psi(\delta \omega,T)\rangle =\frac{1}{\sqrt{2}} \left ( \exp[-\frac{i\delta \omega}{\hbar } \int_{0}^{T}\mu _{max}dt]|\tilde{\psi}_{max}(T)\rangle+\exp[-\frac{i \delta \omega}{\hbar }\int_{0}^{T}\mu _{min}dt]|\tilde{\psi}_{min}(T)\rangle  \right ).
\end{equation}
It is worth noting that what needs to be distinguished is not the input state $|\psi(0)\rangle$ and the output state $|\psi(\delta \omega,T)\rangle$, but the output state $|\psi(\delta \omega,T)\rangle$ with signal and the output state $|\psi(\omega_c,T)\rangle$ without signal, where 
\begin{equation}
	|\psi(\omega_c,T)\rangle =\frac{1}{\sqrt{2}} \left ( \exp[\frac{i\omega_c}{\hbar }  \int_{0}^{T}\mu _{max}dt]|\tilde{\psi}_{max}(T)\rangle+\exp[\frac{i\omega_c}{\hbar } \int_{0}^{T}\mu _{min}dt]|\tilde{\psi}_{min}(T)\rangle  \right ).
\end{equation}
It can define an equivalent Hamiltonian $\hat{G}=\sum_{k}\omega \mu _k(t)| \tilde{\psi}_k(T)  \rangle  \langle \tilde{\psi} _k(T)  |$ between $|\psi(\delta \omega,T)\rangle$ and $|\psi(\omega_c,T)\rangle$, that is
\begin{equation}
	|\psi(\delta \omega,T)\rangle=e^{-\frac{i}{\hbar } \int_{0}^{T} Gdt}|\psi(\omega_c,T)\rangle.
\end{equation}
Obviously, $[\hat{G}(t), \hat{G}\left(t^{\prime}\right)]=0$, it can be equivalently considered that the eigenvector is invariant when using the equivalent Hamiltonian $G$ to obtain information, which goes back to the previous case, and therefore, we can subsequently write the quantum speed limits as
\begin{equation}{\label{QCQSL}}
	\begin{gathered}
		\frac{\omega}{2} \int_{0}^{t_{MT}}(\mu _{max}(t)-\mu _{min}(t))dt \geq \frac{\pi \hbar }{2} \beta (F), \\
		\frac{\omega}{2} \int_{0}^{t_{ML}}(\mu _{max}(t)-\mu _{min}(t))dt \geq \frac{\pi \hbar }{2} \alpha (F).
	\end{gathered}
\end{equation}
It is not hard to find that $t_{MT}\geq t_{ML}$, that is, MTB completely covers MLB, and hence, the time lower bound $\tau$ is a solution to the first inequality of Eq.(\ref{QCQSL}). In addition, the actual evolution process is
\begin{equation}
	F(|\psi(\delta \omega)\rangle,|\psi(\omega_c)\rangle)=\cos^2[\frac{\omega}{2}\int_{0}^{t}(\mu _{max}(t)-\mu _{min}(t))dt ].
\end{equation}
It can be calculated that the actual evolution time $t_{evo}=\tau$. In other words, the quantum speed limit is saturated at this point, minimizing the interrogation time required to obtain a detectable signal.

The above results mean that if we want to reach the quantum speed limit, we need to ensure that the instantaneous eigenvector of the signal Hamiltonian remains unchanged, or equivalently $[\hat{H}(t), \hat{H}\left(t^{\prime}\right)] = 0$, which can be achieved with the help of quantum control techniques. When the input state is an equal-weight superposition of the largest and smallest eigenvectors, the quantum speed limit saturates and the measurement is optimal.
Similarly, as in the case of a time-independent Hamiltonian, if there is insufficient prior information (e.g., if the form of the signal Hamiltonian is unknown), the MLB may offer a tighter estimate. However, this estimate is unsaturated and suboptimal.

\section{Measurement of time-dependent magnetic field signal}
\label{MOTDM}

In this section, we will discuss the temporal resolution limit and the associated constraints with two examples of magnetic field measurements.
Firstly, consider a single mode AC magnetic signal along the Z-axis, the Hamiltonian is
\begin{equation}
	\hat{H}(t)=\mu_{B} B_{0} \cdot \sin(kt) \hat{\sigma}_{z}=\frac{1}{2} \hbar \omega \cdot \sin(kt)  \hat{\sigma}_{z},
\end{equation}
where $\omega=2\mu_{B} B_{0}/\hbar$, which represents the amplitude of the magnetic field, and $k$ is the oscillation frequency of the magnetic field signal, usually $k \geq 0$.

According to the distinguishable condition of quantum states Eq.(\ref{FFODC}), only when the fidelity $F=\cos^2[\frac{\omega }{2}\int_{0}^{t}\sin (kt) dt ]$ between the input and the output state is less than the critical fidelity $F_0=N/(N+1)$, the signal can be considered an effective measurement. It can be calculated that
\begin{equation} \label{KSC}
	k \leq \frac{\omega}{\arccos \left(\sqrt{F_{0}}\right)}.
\end{equation}

This result indicates that for a fixed measurement resource $N$, there exists an upper bound on the oscillation frequency of signal. The signal can only be effectively measured if its frequency is below this threshold. Increasing the measurement resources can enhance this upper bound. For instance, increasing the sampling number $N$ can improve the critical fidelity $F_0$ and thus extend the upper bound. Additionally, utilizing an $M$-body product state or entangled state can accelerate the evolution and relax the constraints. For example, in the case of an $M$-body GHZ entangled state, the upper bound is given by
\begin{equation}
	k \leq \frac{M\omega}{\arccos \left(\sqrt{F_{0}}\right)}.
\end{equation}

\begin{figure*}[!hbt]
	\centering
	\includegraphics[width=14cm]{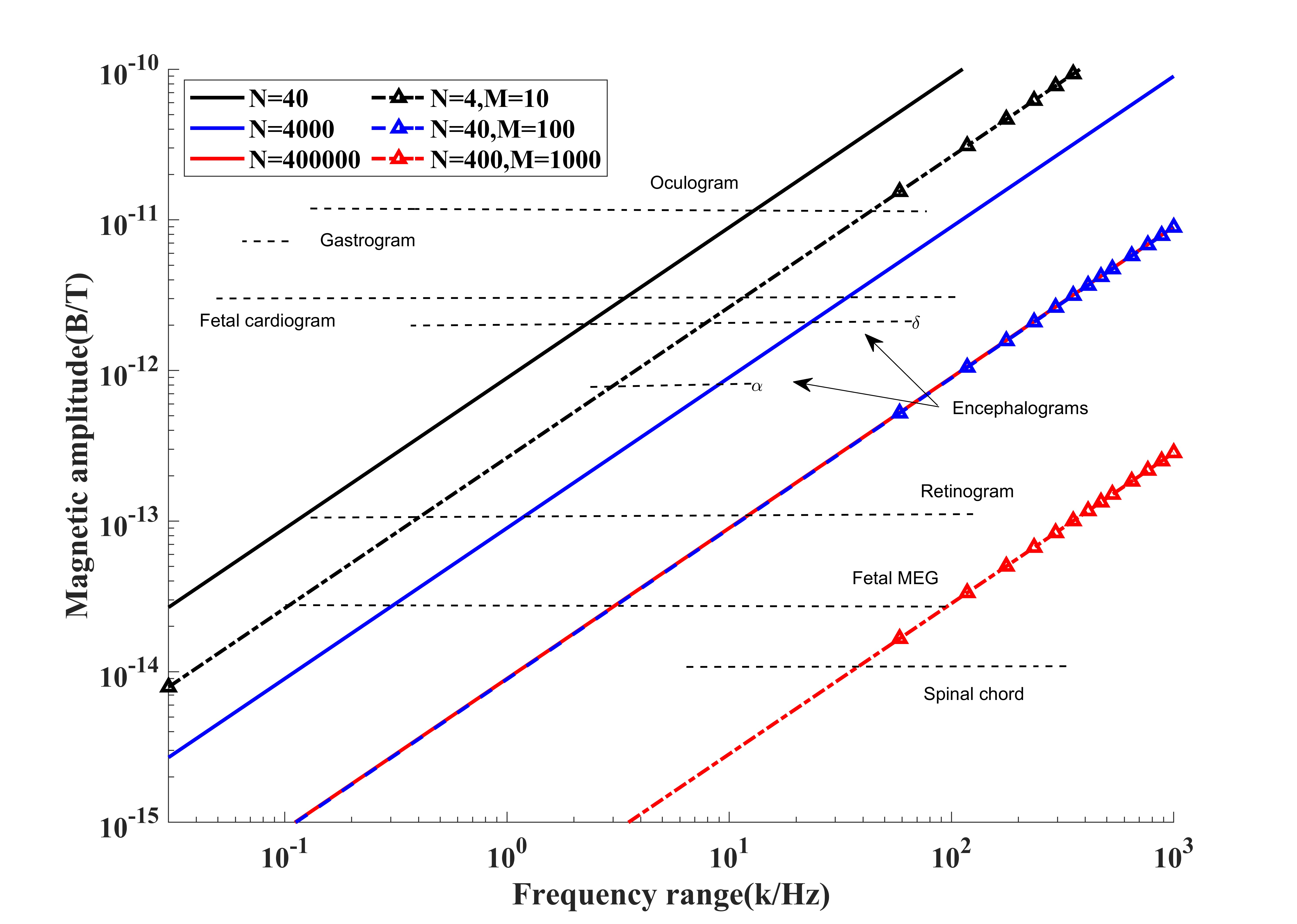}
	\caption{The amplitude and frequency range of biomagnetic fields, as well as the constraints imposed by different resources on magnetic signal, where $N$ represents the number of sampling and $M$ denotes the size of the GHZ entangled state. Only regions above these lines can be effectively measured.}\label{msk}
\end{figure*}

This provides some guidance for the actual application scenario.
Fig.\ref{msk} illustrates the amplitude and frequency range of some biomagnetic signals \cite{RN160}. We calculated the upper bounds of precision and frequency for fixed resources in magnetic measurements. It shows that the faster the magnetic field changes, the more difficult it is to measure, and the more measurement resources are required.

Next, we consider its minimum interrogation time. We already know that the actual evolution time is minimized and the quantum speed limit saturates when the input state is an equal-weight superposition state. Thus, the quantum speed limits are
\begin{equation} \label{TMTL}
	\begin{gathered}
		\int_{0}^{\tau_{M T}}|\sin (k t)| d t=\frac{\pi}{\omega} \cdot \beta\left(F_{0}\right), \\
		\int_{0}^{\tau_{ML}}\sin (k t) d t=\frac{\pi}{\omega} \cdot \alpha\left(F_{0}\right) ,
	\end{gathered}
\end{equation}
the temporal resolution limit is
\begin{equation}
	t_{\min }=\max \left\{\tau_{MT}, \tau_{ML}\right\}=\frac{1}{k} \arccos \left(1-\frac{2 k}{\omega} \arccos \left(\sqrt{F_{0}}\right)\right).
\end{equation}
This conclusion holds for both measurements $\omega$ and $k$; of course, note that Eq.(\ref{KSC}) must be satisfied.

The second example we consider is a rotating magnetic field
\begin{equation}
	\hat{H}(t)=\frac{1}{2} \hbar \omega [\cos (kt)\hat{\sigma}_{x}+\sin(kt) \hat{\sigma}_{z}].
\end{equation}
For this type of Hamiltonian, quantum control techniques must be employed to achieve optimal measurement. Furthermore, the measurement of $\omega$ and $k$ should follow distinct schemes, resulting in different temporal resolutions.
Based on the previous conclusions, for the parameters $\omega$ and $k$ there are
\begin{equation} 
	\begin{gathered}
		\partial _{\omega } \hat{H}(t)=\frac{1}{2} \hbar [\cos (kt)\hat{\sigma}_{x}+\sin(kt) \hat{\sigma}_{z}],\\
		\partial _{k} \hat{H}(t)=\frac{1}{2} \hbar \omega t [-\sin (kt)\hat{\sigma}_{x}+\cos(kt) \hat{\sigma}_{z}].
	\end{gathered}
\end{equation}
Therefore, for the parameter $\omega$, $\mu^{(\omega)} _{max}=\hbar /2 ,\mu^{(\omega)} _{min}=-\hbar/ 2 $, and the corresponding quantum speed limits are
\begin{equation} 
	\begin{gathered}
		t^{(\omega)}_{MT}\ge \tau^{(\omega)}_{MT}= \frac{\pi}{\omega } \cdot \beta (F_0),\\
		t^{(\omega)}_{ML}\ge \tau^{(\omega)}_{ML}= \frac{\pi}{\omega } \cdot \alpha (F_0).
	\end{gathered}
\end{equation}
Similarly, for the parameter $k$, $\mu^{(k)} _{max}=\hbar \omega t /2 ,\mu^{(k)} _{min}=-\hbar\omega t / 2$, and assume that $k=\epsilon \omega$, then
\begin{equation} 
	\begin{gathered}
		t^{(k)}_{MT}\ge \tau^{(k)}_{MT}= \sqrt{\frac{2\pi\cdot\beta (F_0) }{\epsilon \omega^2 }}, \\
		t^{(k)}_{ML}\ge \tau^{(k)}_{ML}= \sqrt{\frac{2\pi\cdot\alpha  (F_0) }{\epsilon \omega^2 }}. 
	\end{gathered}
\end{equation}

Obviously, MTB covers MLB, so the minimum interrogation time is $t_{min}=\tau_{MT}$.
Note that although quantum control techniques are employed here, the minimum interrogation time is only determined by the signal Hamiltonian and the parameter itself.

\section{Discussion}
\label{discussion}

For any type of signal Hamiltonian--whether time-dependent or time-independent, linear or nonlinear--a corresponding lower bound on the interrogation time can be calculated. Violating this bound will inevitably result in the signal being drowned in noise. This lower bound not only constrains the optimal interrogation time and coherence time in quantum sensing but also establishes the fundamental temporal resolution limit for quantum sensing.

The theoretical work in Ref.\cite{RN977} has also explored the impact of of the quantum speed limit on temporal resolution. That study applied both the control rotation sequences and the signal Hamiltonian simultaneously to the quantum state, which reduced the overall time cost. However, as a price, the quantum state is not always in the optimal state, and its sensitivity is subject to the influence of the control rotation angle. In addition, although it mentioned that improving the temporal resolution will affect the signal-to-noise ratio of the measurement results, it did not explicitly provide a lower bound for temporal resolution.
In our work, we consider the control rotation sequences as parts of the state preparation and readout stages. Since these do not involve interaction with the signal itself, they are not considered time critical steps and therefore do not fundamentally influence temporal resolution. Moreover, within our framework, the quantum state always remains in its optimal configuration. This not only maximizes sensitivity but also ensures the saturation of the quantum speed limit, allowing for an accurate calculation of the temporal resolution lower bound. In fact, if the quantum state distinguishable condition (i.e., $SNR \geq 1$) is taken into account, the temporal resolution presented in Ref.\cite{RN977} would not surpass the fundamental temporal resolution limit derived in this paper.

Experimentally, there are no fundamental difficulties in approaching the temporal resolution limit.
The state preparation and readout processes can always be made sufficiently short using high-power control pulses. Alternatively, the quantum state can be prepared in advance using quantum memory, and then perform measurements via delayed-triggering techniques. These techniques hold promise for reaching the fundamental temporal resolution limit in experimental settings. 
It is worth emphasizing that while these methods may not reduce the overall time cost of the entire measurement process, these additional time costs don't impact the temporal resolution of the signal measurement, as the only time critical step is the interrogation stage during which the quantum state interacts with the signal.

\begin{appendices}
	
	\section{Mandelstam–Tamm bound}\label{APPB}
	
	Mandelstam and Tamm \cite{RN296} firstly derived a quantum speed limit under the time-independent Hamiltonian based on the Heisenberg uncertainty relation and the Heisenberg equation of motion, 
	\begin{equation} \label{HUR}
		\Delta \hat{O} \cdot \Delta \hat{H} \geq \frac{1}{2}|\langle\Psi|[\hat{H}, \hat{O}]| \Psi\rangle|,
	\end{equation}
	\begin{equation} \label{HEM}
		i \hbar \frac{d}{d t}\langle\hat{O}\rangle =-\langle\Psi|[\hat{H}, \hat{O}]| \Psi\rangle.
	\end{equation}
	Take the operator $\hat{O}$ as the projection operator of the input state, that is $ \hat{O}=\left|\psi_{0}\right\rangle\left\langle\psi_{0}\right| $, 
	we have
	\begin{equation}
		\begin{gathered}
			\langle\hat{O}\rangle=\left\langle\psi_{t} \mid \psi_{0}\right\rangle\left\langle\psi_{0} \mid \psi_{t}\right\rangle=F\left(\left|\psi_{0}\right\rangle,\left|\psi_{t}\right\rangle\right), \\
			\Delta \hat{O}=\sqrt{F-F^{2}},
		\end{gathered}
	\end{equation}
	substitute into  Eq.(\ref{HUR}) and Eq.(\ref{HEM}) , it can be obtained
	\begin{equation} \label{Hdt}
		\Delta H \cdot d t=\frac{\hbar}{2}\left|\frac{d F}{\sqrt{F-F^{2}}}\right|.
	\end{equation}
	The lower bound of the 
	evolution time \cite{RN296,RN304} can be obtained by directly integrating the  Eq.(\ref{Hdt})
	\begin{equation}\label{GEMT0}
		t_{MT} \geq \frac{\pi \hbar}{2 \Delta H} \cdot \beta(F),
	\end{equation}
	where $\beta(F)=2 \arccos \sqrt{F}/\pi$ , and $0 \leq \beta(F) \leq 1$, which represents the distance between the input state and output state.
	In the case of time-dependent Hamiltonian, the quantum speed limit can be obtained as
	\begin{equation} \label{GEMT1}
		\int_{0}^{t_{MT}} \Delta H(t) d t \geq \frac{\pi \hbar}{2 } \cdot \beta(F).
	\end{equation}
	The lower bound on time is implied by this inequality. However, the computation of time-dependent $ \Delta H(t)$ is not easy and there is generally no analytic expression except in the case of  $\left[\hat{H}(t), \hat{H}\left(t^{\prime}\right)\right] = 0$. Fortunately, this difficulty can be circumvented to some extent for the issues we are concerned with in this paper. Specifically, with the help of quantum control techniques, the equivalent Hamiltonian $\hat{G}=\sum_{k}\omega \mu _k(t)| \tilde{\psi}_k(T)  \rangle  \langle \tilde{\psi} _k(T)  |$ is still satisfied $[\hat{G}(t), \hat{G}\left(t^{\prime}\right)]=0$. Therefore, MTB can also be obtained from
	\begin{equation}\label{GEMT2}
		\int_{0}^{t_{MT}} \Delta G(t) d t \geq \frac{\pi \hbar }{2} \beta (F)
	\end{equation}
	When the input state is an equal-weight superposition of the largest and smallest eigenvectors, the above three equations (i.e.Eq.(\ref{GEMT0},\ref{GEMT1},\ref{GEMT2})) saturate and reach the lowest bound. This is a natural conclusion from a geometric point of view, since the quantum state at this point evolves along a geodesic on the Bloch sphere. At this time, the most information can be obtained, so the measurement is optimal.

	\section{Margolus–Levitin bound}\label{APPC}
	
	Based on the triangle inequality
	\begin{equation} \label{TIE}
		\cos x \geq 1-\frac{2}{\pi}(x+\sin x), \quad x \geq 0
	\end{equation} 
	Margolus and Levitin considered the quantum speed limit when the initial and final states are orthogonal. \cite{RN256}. For the time-independent Hamiltonian, the quantum state evolves from the input state
	\begin{equation}
		\left|\psi_{0}\right\rangle=\sum_{n} c_{n}\left|E_{n}\right\rangle
	\end{equation}
	to the output state
	\begin{equation} \label{FS}
		\left|\psi_{\mathrm{t}}\right\rangle=\sum_{n} c_{n} e^{-i\left(E_{n} t / \hbar\right)}\left|E_{n}\right\rangle,
	\end{equation}
	let
	\begin{equation}
		S(t)=\left\langle\psi_{0} \mid \psi_{t}\right\rangle=\sum_{n}\left|c_{n}\right|^{2} e^{-i\left(E_{n} t / \hbar\right)}.
	\end{equation}
	Thus,
	\begin{equation} \label{RES}
		\begin{split}
			\operatorname{Re}(S)&=\sum_{n}\left|c_{n}\right|^{2} \cos \left(E_{n} t / \hbar\right) \\
			&\geq \sum_{n}\left|c_{n}\right|^{2}\left(1-\frac{2}{\pi}\left(\frac{E_{n} t}{\hbar}+\sin \left(E_{n} t / \hbar\right)\right)\right) \\
			&=1-\frac{2\langle H\rangle t}{\pi \hbar}+\frac{2}{\pi} \operatorname{Im}(S),
		\end{split}
	\end{equation}
	where the triangle inequality Eq.(\ref{TIE}) is used in step 2.
	
	The triangle inequality Eq.(\ref{TIE}) holds if and only if$\quad x \geq 0$,  that is  $E_{n}t/\hbar \geq 0$.
	This requires that all eigenvalues of the quantum state satisfy $E_{n} \geq 0$. In the original paper \cite{RN256}, the authors chose the ground state energy as $E_{g}=0$. When the input quantum state evolves to the corresponding orthogonal state, $S(t)=0$, the Margolus-Levitin inequality can be obtained
	\begin{equation}
		T_{F=0} \geq \frac{\pi \hbar}{2\langle H\rangle}.
	\end{equation}
	However, the condition $E_{g}=0$ is not obvious, to satisfy this condition, we need to add an auxiliary Hamiltonian $-E_{g} \hat{I}$ to the original Hamiltonian $\hat{H}$, so that the complete Hamiltonian is
	\begin{equation} \label{MH}
		\hat{H}^{\prime}=\hat{H}-E_{g} \hat{I}.
	\end{equation}
	Auxiliary Hamiltonian will not affect the evolution of the quantum state, 
	its effects are just to multiply a global phase factor $ e^{-i\left(E_{g} t / \hbar\right)}$, and give a global shift $E_{g}$ to all the eigenvalues.
	The eigenvalues of modified Hamiltonian are $E_{n}^{\prime}=E_{n}-E_{g} \geq 0$, the  ground state energy level is $E_{g}^{\prime}=0$, which satisfies the condition of inequality Eq.(\ref{TIE}). In Eq.(\ref{RES}), replacing $E_{n}$ with $E_{n}^{\prime}$, we can get
	\begin{equation} \label{ML1}
		T_{F=0} \geq \frac{\pi \hbar}{2\langle H^{\prime}\rangle}
	\end{equation}
	or
	\begin{equation} \label{ML2}
		T_{F=0} \geq \frac{\pi \hbar}{2\langle H-E_g\rangle}.
	\end{equation}
	Therefore, more caution should be taken about the Margolus-Levitin inequality. When the expected value of the Hamiltonian is used to measure the quantum speed limit, the Hamiltonian must be the modified Hamiltonian, so the Margolus-Levitin inequality is Eq.(\ref{ML1}). If expressed in terms of the expected value of the original Hamiltonian, it is necessary to consider the value of ground state energy $E_g$ , that is Eq(\ref{ML2}). Notably, the result of Eq. (\ref{ML2}) has been discussed in several references \cite{RN317}.
	
	In quantum mechanics, energy levels are not directly observable; we can only observe the differences between energy levels. Similar to classical mechanics, where different reference planes can be chosen to define potential energy, the definition and selection of energy levels in quantum mechanics also possess some degrees of freedom. Essentially, the modification of Eq. (\ref{MH}) involves re-selecting a reference plane for the energy levels, which uniformly shifts the energy eigenvalues and introduces a global phase factor to the quantum state, without affecting the evolution process. The choice of an energy reference level is not unique. In deriving Eq. (\ref{RES}), we can arbitrarily select an energy reference level $E_r$ as long as $E_n-E_r \geq 0$. The Margolus-Levitin inequality reduces to
	\begin{equation}
		T_{F=0} \geq \frac{\pi \hbar}{2\langle H-E_r\rangle}.
	\end{equation}
	
	If the selected reference level $E_r \leq E_g$, it can be obtained a looser bound than the original Margolus-Levitin bound Eq.(\ref{ML2}). In other words, the Margolus-Levitin inequality is tightest when the ground state energy $E_g$ is chosen as the reference level. The choice of reference level does not affect the evolution of quantum state, and the change of inequality (loose or tight) is just some mathematical tricks. 
	In short, a series of quantum speed limits can be obtained as long as $E_n-E_r \geq 0$. A recent study\cite{RN491} shows that applying a time-reversal operator to the Hamiltonian can obtain a new dual Margolus-Levitin limit, which essentially selects a new reference level.
	
	By extending the above conclusion to any non-orthogonal quantum state, it can be obtained \cite{RN304}
	\begin{equation}
		t \geq \frac{\pi \hbar}{2\langle H-E_r\rangle} \cdot \alpha(F),
	\end{equation}
	where $\alpha(F) \simeq \beta^{2}(F)=\frac{4}{\pi^{2}} \arccos ^{2} \sqrt{F}$, 
	and $0 \leq \alpha(F) \leq 1$, which also represents the distance between the input state and output state.
	It should be noted that $\alpha(F)$ is just a numerical approximation result. It is still an open problem how to use the Margolus-Levitin inequality to describe the quantum speed limit of arbitrary fidelity analytically.
	
	Next, we consider two time-dependent cases. If $\left[\hat{H}(t), \hat{H}\left(t^{\prime}\right)\right] = 0$, we can get
	\begin{equation}
		\left|\psi_{t}\right\rangle=\sum_{n} c_{n} e^{-i \int_{0}^{t}E_{n}(t)dt / \hbar}\left|E_{n}\right\rangle.
	\end{equation}
	The ground state is chosen as the reference level, that is $E_r=E_{min}(t)$, MLB can be written as
	\begin{equation}\label{MLB1}
		\int_{0}^{t_{ML}} (\left \langle H(t) \right \rangle-E_{min}(t))dt \geq \frac{\pi \hbar }{2} \cdot \alpha  (F)
	\end{equation}
	Similarly, even if $\left[\hat{H}(t), \hat{H}\left(t^{\prime}\right)\right] \ne 0$, with the help of quantum control techniques, the evolution of quantum states can be written as
	\begin{equation}
		|\psi(\delta \omega,T)\rangle=e^{-\frac{i}{\hbar } \int_{0}^{T} Gdt}|\psi(g_c,T)\rangle=\sum_{n} c_{n} e^{-i \int_{0}^{t} \omega \mu_{n}(t)dt / \hbar}   |\tilde{\psi}_{n}(T)\rangle
	\end{equation}
	so MLB is
	\begin{equation}\label{MLB2}
		\int_{0}^{t_{ML}} (\left \langle G \right \rangle-\mu_{min}(t))dt \geq \frac { \pi \hbar }{2} \cdot \alpha  (F),
	\end{equation}
	which is actually the same as Eq.(\ref{MLB1}).
	
\end{appendices}

\backmatter

\bmhead{Acknowledgements}

Not applicable.

\bmhead{Authors' contributions}

All authors made a significant contribution to the work and were involved in interpreting the results and writing the manuscript.

\bmhead{Funding}
This work is supported by NSAF (Grant No. U2130205) and Innovational Fund for Scientific and Technological Personnel of Hainan Province (Grant No. KJRC2023B11).

\bmhead{Data availability}

Not applicable.

\section*{Declarations}

\bmhead{Ethics approval and consent to participate}

Not applicable.

\bmhead{Consent for publication}

Not applicable.

\bmhead{Competing interests}

The authors declare that they have no competing interests.

\end{document}